\documentclass[12pt]{article}
\usepackage{latexsym,epsfig,graphicx,a4wide,amssymb}

{\rm }

\def\be{\begin{equation}}
 \def\ee{\end{equation}}
 \def\bea{\begin{eqnarray}}
 \def\eea{\end{eqnarray}}
\newcommand{\fr}{\frac}

\def\2{\frac{1}{2}}
\def\4{\frac{1}{4}}

\def\PLA#1{{Phys.\ Lett.\ A \bf #1}}
\def\PR#1{{Phys.\ Rev.\ D \bf #1}}
\def\PRL#1{{Phys.\ Rev.\ Lett.\ \bf #1}}
\def\cm#1{{Commun.\ Math.\ Phys.\ \bf #1}}

\def\cqg#1{{Class.\ Quant.\ Grav.\ \bf #1}}

\catcode`\@=11

\@addtoreset{equation}{section}

\def\@normalsize{\@setsize\normalsize{15pt}\xiipt\@xiipt
\abovedisplayskip 14pt plus3pt minus3pt%
\belowdisplayskip \abovedisplayskip
\abovedisplayshortskip  \z@ plus3pt%
\belowdisplayshortskip  7pt plus3.5pt minus0pt}
\def\small{\@setsize\small{13.6pt}\xipt\@xipt
\abovedisplayskip 13pt plus3pt minus3pt%
\belowdisplayskip \abovedisplayskip
\abovedisplayshortskip  \z@ plus3pt%
\belowdisplayshortskip  7pt plus3.5pt minus0pt
\def\@listi{\parsep 4.5pt plus 2pt minus 1pt
            \itemsep \parsep
            \topsep 9pt plus 3pt minus 3pt}}

\def\underline#1{\relax\ifmmode\@@underline#1\else
        $\@@underline{\hbox{#1}}$\relax\fi}
\@twosidetrue \relax

\catcode`@=12

\catcode`\@=11

\def\section{\@startsection{section}{1}{\z@}{3.5ex plus 1ex minus
   .2ex}{2.3ex plus .2ex}{\large\bf}}
\def\ps@headings{\def\@oddfoot{}\def\@evenfoot{}
\def\@oddhead{\hbox{}\hfill
        \makebox[.5\textwidth]{\raggedright\ignorespaces --\thepage{}--
        \hfill }}
\def\@evenhead{\@oddhead}
\def\subsectionmark##1{\markboth{##1}{}}
}

\ps@headings

\catcode`\@=12

\begin{document}

\begin{titlepage}
%
%

%\rightline{May 2012}

%

\begin{centering}
\vspace{1cm}
%i
{\Large {\bf Scalar Hair from a Derivative Coupling of a Scalar
Field to the Einstein Tensor}}\\

\vspace{1.5cm}

 {\bf Theodoros Kolyvaris $^{\dagger}$}, {\bf George Koutsoumbas $^{\sharp}$},\\
{\bf Eleftherios Papantonopoulos $^{*}$} \\
 \vspace{.2in}
 Department of Physics, National Technical University of
Athens, \\
Zografou Campus GR 157 73, Athens, Greece \\
\vspace{.2in}
 {\bf George Siopsis $^{\flat}$}
\vspace{.2in}

 Department of Physics and Astronomy, The
University of Tennessee,\\ Knoxville, TN 37996 - 1200, USA
 \\
\vspace{3mm}

\end{centering}
\vspace{1.5cm}

\begin{abstract}

We consider a gravitating system of vanishing cosmological
constant consisting of an electromagnetic field and a scalar field
coupled to the Einstein tensor. A Reissner-Nordstr\"om black hole
undergoes a second-order phase transition  to a hairy black hole
of generally anisotropic hair at a certain critical temperature
which we compute. The no-hair theorem is evaded due to the
coupling between the scalar field and the Einstein tensor. Within
a first order perturbative approach we calculate explicitly the
properties of a hairy black hole configuration near the critical
temperature and show that it is energetically favorable over the
corresponding Reissner-Nordstr\"om black hole.

\end{abstract}

\vspace{3.5cm}

\begin{flushleft}
%$^{\dagger}~~$ e-mail address: <a href="/src/compose.php?send_to=teokolyv%40central.ntua.gr">teokolyv@central.ntua.gr</a> \\
%$^{\sharp}~~$ e-mail address: <a href="/src/compose.php?send_to=kutsubas%40central.ntua.gr">kutsubas@central.ntua.gr</a> \\
%$^{*} ~~$ e-mail address: <a href="/src/compose.php?send_to=lpapa%40central.ntua.gr">lpapa@central.ntua.gr</a> \\
%$ ^{\flat}~~$ e-mail address: <a href="/src/compose.php?send_to=siopsis%40tennessee.edu">siopsis@tennessee.edu</a>
$^{\dagger}~~$ e-mail address: teokolyv@central.ntua.gr \\
$^{\sharp}~~$ e-mail address: kutsubas@central.ntua.gr \\
$^{*} ~~$ e-mail address: lpapa@central.ntua.gr \\
$ ^{\flat}~~$ e-mail address: siopsis@tennessee.edu

\end{flushleft}
\end{titlepage}

\section{Introduction}

The ``no-hair" theorems are powerful tools in studying black hole
solutions of Einstein gravity coupled with matter. These
``no-hair" theorems describe the existence and stability of
four-dimensional asymptotically flat black holes coupled to an
electromagnetic field or in vacuum.  In the case of a minimally
coupled scalar field in asymptotically flat spacetime the ``no-hair"
theorems were proven imposing conditions on the form of the
self-interaction potential~\cite{nohairtheo}. These theorems were
also generalized to non-minimally coupled scalar
fields~\cite{Mayo}.

For asymptotically flat spacetime, a four-dimensional black hole
coupled to a scalar field with a zero self-interaction potential
is known~\cite{BBMB}. However, the scalar field diverges on the
event horizon and, furthermore, the solution is unstable
\cite{bronnikov}, so there is no violation of the ``no-hair"
theorems.   In the case of a positive cosmological constant with a
minimally coupled scalar field with a self-interaction potential,
black hole solutions were found in~\cite{Zloshchastiev:2004ny} and
also a numerical solution was presented in~\cite{Torii:1998ir},
but it was unstable. If the scalar field is non-minimally coupled,
a solution exists with a quartic self-interaction
potential~\cite{martinez}, but it was shown to be
unstable~\cite{phil,dotti}.

In the case of a negative cosmological constant, stable solutions
were found numerically for spherical
geometries~\cite{Torii:2001pg, Winstanley:2002jt} and an exact
solution in asymptotically AdS space with hyperbolic geometry was
presented in~\cite{Martinez:2004nb} and generalized later to
include charge~\cite{Martinez:2005di} and further generalized to
non-conformal solutions \cite{Kolyvaris:2009pc}. This solution is
perturbatively stable for negative mass and may develop
instabilities for positive mass~\cite{papa1}. The thermodynamics
of this solution was studied in~\cite{Martinez:2004nb} where it
was shown that there is a second order phase transition of the
hairy black hole to a pure topological black hole without hair.
The analytical and numerical calculation of the quasi-normal modes
of scalar, electromagnetic and tensor perturbations of these black
holes confirmed this behaviour~\cite{papa2}. Recently, a new exact
solution of a charged C-metric conformally coupled to a scalar
field was presented in~\cite{kolyvaris, anabalon}. A
Schwarzschild-AdS black hole in five-dimensions coupled to a
scalar field was discussed in~\cite{Farakos:2009fx}, while
dilatonic black hole solutions with a Gauss-Bonnet term in various
dimensions were discussed in~\cite{Ohta:2009pe}.

Recently scalar-tensor theories  with nonminimal couplings between
derivatives of a scalar field and curvature were studied.  The
most general gravity Lagrangian linear in the curvature scalar R,
quadratic in the scalar field $\phi$, and containing terms with four derivatives
was considered in \cite{Amendola:1993uh}. It was shown that this
theory cannot be recast into Einsteinian form by a conformal
rescaling. It was further shown that  without considering any
effective potential, an effective cosmological constant and then
an inflationary phase can be generated.

Subsequently it was found \cite{Sushkov:2009hk}  that the equation
of motion for the scalar field can be reduced to a second-order
differential equation when the scalar field is kinetically coupled to the
Einstein tensor. Then the cosmological evolution of the scalar
field coupled to the Einstein tensor was considered and it was shown
that the universe at early stages  has a quasi-de Sitter behaviour
corresponding to a cosmological constant proportional to the
inverse of the coupling of the scalar field to the Einstein tensor.
These properties of the derivative coupling of the scalar field to
curvature had triggered the interest of the study of the
cosmological implications of this new type of scalar-tensor theory
\cite{Gao:2010vr,Granda:2009fh,Saridakis:2010mf,Germani:2010gm}.
Also local black hole solutions were discussed in
\cite{Germani:2011bc}.

The dynamical evolution of a scalar field coupled to the Einstein
tensor in the background of a Reissner-Nordstr\"om black hole
was studied in \cite{Chen:2010qf}. By calculating the quasinormal
spectrum of scalar perturbations it was found that for weak
coupling of the scalar field to the Einstein tensor and for small
angular momentum the effective potential outside the horizon of
the  black hole is always positive  indicating that the background
black hole is stable for a weaker coupling. However, for higher
angular momentum and as the coupling constant  gets larger
than a critical value, the effective potential develops a negative
gap near the black hole horizon indicating an instability of the
black hole background.

The previous discussion indicates that the presence of the
derivative coupling of a scalar field to the Einstein tensor on
cosmological or black hole backgrounds generates an effect
similar to the presence of an effective cosmological constant. In
this paper we  investigate this effect further. We consider a
spherically symmetric Reissner-Nordstr\"om black hole
and perturb this background by introducing a derivative
coupling of a scalar field to the Einstein tensor. We  show that in
this gravitating system there exists a critical temperature in
which the system undergoes a second-order phase transition to an
anisotropic hairy black hole configuration and the scalar field is
regular on the horizon. This ``Einstein hair" is the result of
evading the no-hair theorem thanks to the presence of the derivative
coupling of the scalar field to the Einstein tensor.

The coupled dynamical system of Einstein-Maxwell-Klein-Gordon
equations is a highly non-linear system of equations for which even
a numerical solution appears beyond reach. To solve the field equations, we expand
the fields around a Reissner-Nordstr\"om black hole solution
and pertubatively determine the critical temperature.  By solving the first-order equations
numerically near the critical temperature, we study the behaviour
of the hairy black hole solution. We calculate
 the temperature of the new hairy black hole and compare it  with
the corresponding temperature of a Reissner-Nordstr\"om black
hole of the same charge. We find that above the critical temperature the
Reissner-Nordstr\"om black hole is unstable and by calculating
the free energies we show that the new hairy black hole
configuration is  energetically favorable over the corresponding
Reissner-Nordstr\"om black hole.

The paper is organized as follows. In Sec.\ \ref{sec:1} we set up
the field equations and outline  our solution. In Sec.\
\ref{sec:2} we find the zeroth order solution and calculate the
critical temperature near which the Reissner-Nordstr\"om black
hole may become unstable and develop hair. In Sec.\ \ref{sec:3} we
find the first-order solutions to the system of
Einstein-Maxwell-Klein-Gordon equations which are hairy black
holes near the critical temperature. In Sec.\ \ref{sec:5} we
discuss the thermodynamic stability of our first-order hairy
solution. In Sec. \ref{sec:4} we discuss the validitiy of our perturbative expansion. Finally, in Sec.\ \ref{sec:6} we conclude.

\section{The field equations}
\label{sec:1}

%\section{Static Hairy Black Hole Solutions}

Consider the Lagrangian density \be \mathcal{L} = \frac{R}{16\pi
G} - \frac{1}{4} F_{\mu\nu} F^{\mu\nu} - (g^{\mu\nu} + \kappa
G^{\mu\nu} ) D_\mu \varphi (D_\nu \varphi)^* - m^2 |\varphi|^2~,
\label{lagdens} \ee where $D_\mu = \partial_\mu - ie A_\mu$ and
 $e$, $m$ the charge and mass of the scalar
field and $\kappa $ the coupling of the scalar field to Einstein
tensor, of dimension length squared.

In this paper, we shall concentrate on the case of a massless and chargeless scalar field, setting
\be\label{eqpar} m =0 \ \ , \ \ \ \ e = 0~, \ee
and leave the study of the general case to future work. Consequently, the scalar field $\varphi$ is {\em real}.

With the choice of parameters (\ref{eqpar}), the field equations
resulting from the Lagrangian (\ref{lagdens}) are the Einstein
equations \be R_{\mu\nu} = 8\pi G T_{\mu\nu}~,\label{eineqfull}
\ee the Maxwell equations \be \nabla^\mu F_{\mu\nu} = 0 \ , \ \
F_{\mu\nu} =
\partial_\mu A_\nu - \partial_\nu A_\mu~,\label{maxfull} \ee and the Klein-Gordon
equation for the scalar field, \be \frac{1}{\sqrt{-g}}
\partial_\mu \left[ \sqrt{-g} (g^{\mu\nu} + \kappa G^{\mu\nu} )
\partial_\nu \varphi \right] = 0~.\label{kleingordon} \ee
The stress-energy tensor receives three contributions, \be
T_{\mu\nu} = T_{\mu\nu}^{(EM)} + T_{\mu\nu}^{(\varphi)} + \kappa
\Theta_{\mu\nu}~, \label{stesenerg}\ee where $T_{\mu\nu}^{(EM)}$
is the electromagnetic stress-energy tensor,
$T_{\mu\nu}^{(\varphi)}$ is the standard scalar field
contribution, \be\label{scalarstess} T_{\mu\nu}^{(\varphi)} =
\partial_\mu \varphi
\partial_\nu\varphi - \frac{1}{2} g_{\mu\nu} g^{\alpha\beta} \partial_\alpha\varphi
\partial_\beta \varphi~, \ee
and $\Theta_{\mu\nu}$ is an extra matter source resulting from the derivative coupling of the scalar field to the Einstein tensor,
\bea \Theta_{\mu\nu}&=&
-\fr{1}{2}\nabla_\mu\varphi\nabla_\nu\varphi R
-\fr{1}{2}(\nabla\varphi)^2 G_{\mu\nu} +
\fr{1}{2}\nabla_\mu\nabla_\nu(\nabla\varphi)^2
-\fr{1}{2}g_{\mu\nu}\square(\nabla\varphi)^2 \nonumber\\
& & -\fr{1}{2}g_{\mu\nu}\nabla_\alpha\varphi\nabla_\beta\varphi
R^{\alpha\beta} + 2\nabla_\alpha\varphi \nabla_{(\mu}\varphi
R_{\nu)}^\alpha+\fr{1}{2}\square(\nabla_\mu\varphi\nabla_\nu\varphi)
-\nabla_\alpha\nabla_{(\mu}(\nabla_{\nu)}\varphi\nabla^\alpha\varphi)
\nonumber\\
& & +\fr{1}{2}
g_{\mu\nu}\nabla_\alpha\nabla_\beta(\nabla^\alpha\varphi\nabla^\beta\varphi)~.
 \label{eqtheta}\eea
The field equations have well-known solutions for $\varphi =0$, the Reissner-Nordstr\"om black holes.
We are interested in finding {\em hairy} solutions, with $\varphi \ne 0$. The no-hair theorem will be evaded thanks to the coupling of the scalar field to the Einstein tensor (``Einstein hair'').

To solve the non-linear field equations, we shall expand around a Reissner-Nordstr\"om black hole.
It should be pointed out that not all Reissner-Nordstr\"om black holes can be starting points of such a perturbative expansion.
We will show that for a given charge of the black hole, there is generally a unique mass for which the Reissner-Nordstr\"om black hole can be a starting point. The corresponding temperature is then expected to be a \emph{critical temperature} at which a phase transition occurs from a Reissner-Nordstr\"om black hole to a hairy one.

Introducing the (small) order parameter $\varepsilon$, we therefore set
\bea
\varphi &=& \varepsilon\left( \varphi^{(0)} + \varepsilon^2 \varphi^{(1)} + \dots \right) \nonumber\\
g_{\mu\nu} &=& g_{\mu\nu}^{(0)} + \varepsilon^2 g_{\mu\nu}^{(1)} + \dots \nonumber\\
A_\mu &=& A_\mu^{(0)} + \varepsilon^2 A_\mu^{(1)} + \dots
\label{pertfull}\eea and solve the field equations perturbatively.

To find a static solution of the field equations, it is convenient to consider the metric {\em
ansatz} \footnote{This is a form of the  Lewis-Papapetrou metric
written in isotropic coordinates in which all the metric functions
$l$, $\alpha$ and $\beta$ depend only on $r$ and $\theta $
\cite{Kleihaus:1997ws}.}
\begin{equation}\label{eq1ag} ds^2 = -e^{-\alpha} dt^2 + l^2 e^\alpha \left[ e^{-\beta} (dr^2 + r^2
d\theta^2) + r^2\sin^2\theta d\phi^2 \right] ~, \end{equation}
and electromagnetic potential
\be A_t = A(r,\theta) \ \ , \ \ \ \ \vec A = \vec 0 ~. \ee
%and set
%\be f = e^{-\alpha} \ , \ \ g = e^{-\beta}~. \ee
The metric functions are expanded as
\bea l &=& l_0 + \varepsilon^2 l_1 + \dots \nonumber\\
\alpha &=& \alpha_0 + \varepsilon^2 \alpha_1 + \dots \nonumber\\
\beta &=& \beta_0 + \varepsilon^2 \beta_1 + \dots~, \eea
and the electrostatic potential as
\be A = A_0 + \varepsilon^2 A_1 + \dots ~.\ee
The components of the electromagnetic stress-energy tensor are
%Then the $T_{\mu\nu}^{(EM)}$ is given by
\begin{eqnarray}
T_t^{(EM)t} = - T_\phi^{(EM)\phi} &=& \frac{1}{2r^2l^2 g} \left[ r^2 (\partial_r
A)^2 + (\partial_\theta A)^2 \right]~, \nonumber\\
T_r^{(EM)r} = - T_\theta^{(EM)\theta} &=& \frac{1}{2r^2l^2 g} \left[ r^2
(\partial_r A)^2 - (\partial_\theta A)^2 \right]~, \nonumber\\
T_r^{(EM)\theta} &=& \frac{1}{l^2 g} \partial_r A \partial_\theta
A~. \label{3str}
\end{eqnarray}
%For  the simplest special case $\ell =1,\ m=0,\ e=0$ we are
%considering in this work, the scalar field is of the form
%\[ \varphi = \varphi (z,\theta) = \mathcal{Z}(z) \cos\theta~. \]
The Einstein equations reduce to the convenient subset
\begin{eqnarray} R_t^t +R_\phi^\phi &=& 8\pi G \left( T_t^t + T_\phi^\phi - T
\right) \label{eq1n}~, \\
 R_t^t &=& 8\pi G \left( T_t^t - \frac{1}{2} T \right)  \label{eq2n}~, \\
R_r^r - R_\theta^\theta  &=& 8\pi G \left( T_r^r -T_\theta^\theta \right)
\label{eq3n}~, \end{eqnarray} where
\be T = T_\mu^\mu = -(\partial\varphi)^2 ~,\ee
to be solved together with the Maxwell equation (Gauss's Law), \be
\nabla_\mu \nabla^\mu A = 0~. \label{eq6n}\ee
and the Klein-Gordon equation (\ref{kleingordon}).

The Hawking temperature of the black hole solution can also be expanded as
\be T = T_0 + \varepsilon^2 T_1 + \dots ~. \ee
When $\varepsilon = 0$, the temperature is $T=T_0$. This is a {\em critical temperature} at which a second-order phase transition occurs to a hairy black hole, if the latter solution exists and is energetically favorable. For a small order parameter,
\be \varepsilon \sim |T - T_0|^{1/2} ~, \ee
so our expansion can be viewed as an expansion around the critical temperature.

Next, we proceed with the perturbative solution of the field equations.

%\subsection{Reissner-Nordstr{\o}m black holes}

\section{Zeroth order solution}
\label{sec:2}

At zeroth order we obtain the Einstein-Maxwell equations
\be\label{eq3.1} R_{\mu\nu}^{(0)} = 8\pi G T_{\mu\nu}^{(EM)(0)} \ , \ \ \nabla^\mu F_{\mu\nu}^{(0)} = 0~, \ee
as well as the Klein-Gordon equation
\be\label{eq3.2} \frac{1}{\sqrt{-g^{(0)}}} \partial_\mu \left[ \sqrt{-g^{(0)}} (g^{(0)\mu\nu} + \kappa R^{(0)\mu\nu} )
\partial_\nu \varphi^{(0)} \right] = 0~,\label{kleingordon0} \ee
where we used $G_{\mu\nu}^{(0)} = R_{\mu\nu}^{(0)}$ for the
solution of (\ref{eq3.1}). Thus at this order,
the Einstein-Maxwell equations decouple (effectively, $\varepsilon =0$), and the solution
$g_{\mu\nu}^{(0)}$,  $A_\mu^{(0)}$ has no hair (e.g., a
Reissner-Nordstr\"om black hole).

Note the Klein-Gordon equation for $\varphi^{(0)}$ in the
background $(g_{\mu\nu}^{(0)}\ ,\ A_\mu^{(0)})$ constrains the
background, because not all solutions of the Einstein-Maxwell
equations lead to regular functions $\varphi^{(0)}$. The demand of regularity results
in relations among the parameters of the Reissner-Nordstr\"om solution. Viewed as a
thermodynamic system, the critical temperature $T_0$ is also
constrained by demanding regularity of $\varphi^{(0)}$.

The Einstein-Maxwell
equations at zeroth order (\ref{eq3.1}) have as a solution the
Reissner-Nordstr\"om black hole. In Lewis-Papapetrou coordinates,
\begin{equation} l_0 = 1 - \frac{\mu^2}{r^2} \ , \ \ l_0 e^{\alpha_0/2} = 1
+ \frac{\mu^2}{r^2} + \frac{2\mu\coth B}{r} \ , \ \ \beta_0 =0~,
\label{eq2-3}\end{equation} and the electrostatic potential is
\begin{equation} A_t^{(0)} = \Phi- \frac{Q}{rl_0} e^{-\alpha_0/2} \ \ , \ \ \ \ Q =
\frac{2\mu}{\sqrt{G}\, \sinh B} \ , \ \
\Phi = \frac{e^{-B}}{\sqrt{G}}~, \end{equation} where $Q$ is the
 charge of the black hole, and  $\Phi$ is the value of the
potential as $r\to\infty$, fixed by the requirement that $A_\mu
A^\mu$ be finite at the horizon.

%{\bf What the second referee wants for $\Phi$?}

To put it in more familiar terms, performing the coordinate
transformation
\[ \rho = l_0r e^{\alpha_0/2} = r + \frac{\mu^2}{r} + 2\mu\coth B~, \]
the metric becomes \be ds^2 = - f(\rho) dt^2 +
\frac{d\rho^2}{f(\rho)} + \rho^2 d\Omega^2 \ \ , \ \ \ \ f(\rho) = 1 -
\left( 1 + \frac{G Q^2}{\rho_+^2} \right) \frac{\rho_+}{\rho} +
\frac{G Q^2}{\rho^2}~, \label{fform}\ee showing that the horizon
is at $\rho=\rho_+$~,
\[ \rho_+ = Q \sqrt{ G} \ e^B~. \]
The electrostatic potential reads $A_t^{(0)} = \frac{Q}{\rho}$,
confirming that $Q$ is the charge of the hole. Its temperature is
\be\label{eq2-5a} T = \frac{f'(\rho_+)}{4\pi} = \frac{1}{4\pi
\rho_+} \left( 1 - \frac{G Q^2}{\rho_+^2} \right) = \frac{e^{-2B}
\sinh^2 B}{4\pi\mu}~. \ee The mass, entropy, and potential are,
respectively, \be\label{eq2-5b} M = \frac{2\mu}{G} \coth B \ \ , \
\ \ \ S = \frac{4\pi \mu^2}{G} \frac{e^{2B}}{\sinh^2 B} \ \ , \ \
\ \ \Phi = \frac{e^{-B}}{\sqrt{G}} ~. \ee From these, or using the
Euclidean action, we deduce the Gibbs free energy
\be\label{eqRNGibbs} F = M -TS -Q\Phi = \frac{\mu}{G}~. \ee

%\subsection{The critical temperature}

Additionally, the scalar field obeys the Klein-Gordon equation
(\ref{eq3.2}). For $\kappa =0$, there is no regular solution, which is a
consequence of the no-hair theorem \cite{nohairtheo,Mayo}.
However, for $\kappa\ne 0$, we obtain regular static solutions for
certain values of the parameters $\mu$ and $B$. In
\cite{Chen:2010qf}  the Klein-Gordon equation (\ref{eq3.2}) was
solved numerically in the background of the Reissner-Nordstr\"om
black hole. The corresponding temperature of the
Reissner-Nordstr\"om black hole $T_0$ is the critical
temperature, near which the black hole may become unstable and
develop hair, evading the no-hair theorem.

To calculate the critical temperature, it is convenient to
introduce the coordinate \be z = \frac{\mu}{r} \label{eqz}~, \ee
so that the boundary is at $z=0$ and the horizon at $z=1$.

We look for regular solutions of the Klein-Gordon equation
(\ref{eq3.2}) in the interval $[0,1]$. Such solutions where
discussed in \cite{Chen:2010qf} where an instability was found
in the quasinormal frequencies for multipole number $\ell \ne
0$. To capture such an effect we consider a non-spherical {\em
ansatz}
% Let $\varphi = \frac{1}{\sqrt{2}}\Psi e^{ie\vartheta}$, where $\Psi$ and
%$\vartheta$ are real fields.
\be\label{eqphi0} \varphi^{(0)} = \mathcal{Z} (z) P_\ell
(\cos\theta)~. \ee After some straightforward algebra, we obtain
\be\label{waveeq} \frac{z^2}{l_0} \left( l_0(1+\kappa\mathcal{R})
\mathcal{Z}' \right)' - (1-\kappa\mathcal{R})
\ell(\ell+1)\mathcal{Z} = 0~, \ee where prime denotes
differentiation with respect to $z$ and we defined
%introduced the dimensionless quantities
%
% \[ \hat\kappa = \frac{\kappa}{\mu^2} \ , \ \ \hat{R} = \mu^2 R_t^t = \mu^2 R_r^r =
% - \mu^2 R_\theta^\theta \ , \ \ \hat{m} = \mu m \ , \ \ \hat{A}_t = \mu A_t \]
\be\label{eqcalR} \mathcal{R} = {R^{(0)}}_t^t = {R^{(0)}}_z^z =  - {R^{(0)}}_\theta^\theta =
\frac{8 \sinh^2 B}{\mu^2 } \frac{z^4}{[(1+z^2)\sinh B + 2z \cosh B ]^4}~. \ee
The Klein-Gordon equation (\ref{eq3.2}) does not have a
solution with a regular scalar field at the horizon for $\ell = 0$.
This means that scalar hair, if it exists, is
anisotropic.\footnote{We should point out that spherically symmetric solutions of the Klein-Gordon
equation resulting from the Lagrangian density (\ref{lagdens}) may still exist
in the case of non-vanishing mass and charge of the scalar field. This more general case is under
investigation. } We shall  concentrate on the case of a dipole, \be\label{eqell1} \ell
=1~. \ee
% and leave the study of the general case to future work.
%The wave equation becomes \be\label{waveeq} \frac{z^2}{1-z^2}
%\left( (1-z^2)(1+\kappa\mathcal{R}) \mathcal{Z}' \right)' -
%2(1-\kappa\mathcal{R} ) \mathcal{Z} = 0~. \ee
At the boundary we
obtain $z^2 \mathcal{Z}'' - \ell (\ell +1) \mathcal{Z} \approx 0$,
therefore
\[ \mathcal{Z}\sim z^{\ell +1}~. \]
At the horizon we have $((1-z)\mathcal{Z}')'\approx 0$, therefore
$\mathcal{Z}' \sim 0$ or $\mathcal{Z}'\sim \frac{1}{1-z}$. For
each value of the charge $Q$, there should be a unique combination
of parameters $\mu$ and $B$ that gives a convergent $\mathcal{Z}'$
(and therefore a finite $\mathcal{Z}$) at the horizon. For given
$\kappa$, we will determine numerically pairs of $\mu$ and $B$
from (\ref{waveeq}) and their values will be used to determine
$T_0$ as a function of the charge $Q$ using (\ref{eq2-5a}).

%In this section we will numerically solve the scalar wave equation
%(\ref{waveeq}) and determine the metric functions $\alpha$ and
%$\beta$ and the electric potential $A$.
To solve  the scalar wave equation (\ref{waveeq}) numerically, we
use the  variational method. In particular we use an expansion
around the horizon \be \mathcal{Z}(z) = z^2 W(z)= z^2
\sum_{n=0}^{\infty} W_n (1-z)^n~. \ee It turns out that for a
better than 10\% precision, one needs to keep at least 10 terms in the
expansion.

%{\bf More details as requested by the second referee}

%If we substitute
%$\mathcal{Z}(z)$ in the equation (\ref{waveeq}), we find
%that it is of order $(1-z)^8.$ As implied in the expansion, ten
%terms were needed to get this precision.

\begin{figure}
\begin{center}
\includegraphics[scale=0.9,angle=0]{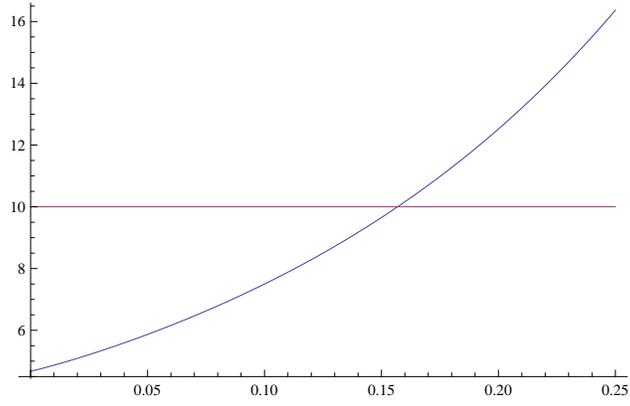}
\end{center}
\caption {Graphical solution of (\ref{eqkappa}) {\em vs.}\ $B$, showing that
$B=0.158$.} \label{kappa}
\end{figure}

\begin{figure}
\begin{center}
\includegraphics[scale=0.9,angle=0]{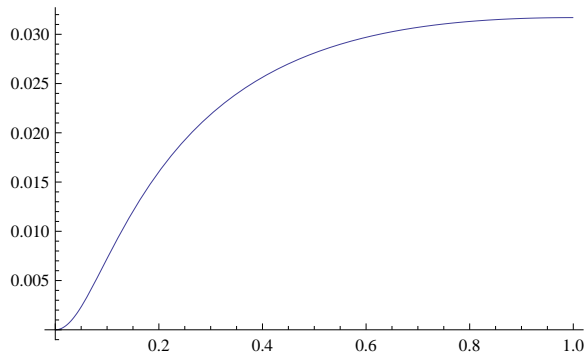}
\end{center}
\caption {The function ${\cal Z}(z)$ for the scalar field versus
$z$ for the parameters of figure \ref{kappa}.} \label{R}
\end{figure}

The coefficients $W_n$ depend on $\kappa$ and $B.$ Our aim is to
find a unique $\kappa$ which will give finite $\mathcal{Z}(z)$ and
$\mathcal{Z}(z)^\prime$ at the horizon $z=1.$ One may rearrange
eq.\ (\ref{waveeq}) to get \be\label{eqkappa} \kappa =
\fr{\int_0^1 dz (1-z^2)[2 \mathcal{Z}(z)^2/z^2+\mathcal{Z}^{\prime
2}(z)]}{\int_0^1 dz (1-z^2){\cal R}(z) [2
\mathcal{Z}(z)^2/z^2-\mathcal{Z}^{\prime 2}(z)]}~, \ee which can
be solved graphically. For the choice \be \kappa = 10 \ , \ \ Q =
9.5~, \label{values}\ee in units in which $8\pi G =1$, we obtain
$\mu = 0.15$, $B = 0.158$ (see figure \ref{kappa}), which gives
the critical temperature \be T_0 \approx 0.01~. \ee In figure
\ref{R}, we depict the scalar function $\mathcal{Z} (z)$ for the
chosen values of the parameters, normalized so that $\mathcal{Z}
(1) \sim T_0$ (the scalar field and the temperature have the same
dimension). The normalization is arbitrary, since the wave
equation is linear.

%{\bf How the normalization effects the results?}

%, but we have chosen a
%small overall normalization that can be used for a perturbative
%expansion near the critical temperature.
The function
$\mathcal{Z}(z)$ is regular in the entire range outside the horizon, as desired.

\begin{figure}
\begin{center}
\includegraphics[scale=0.5,angle=-90]{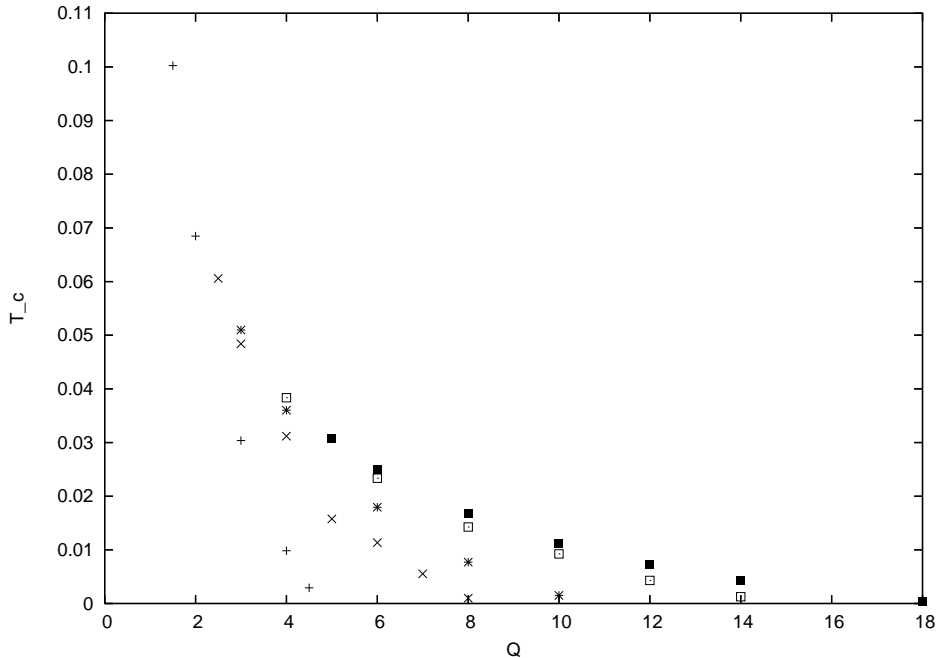}
\end{center}
\caption {The critical temperature $T_0$ \emph{vs.}\ the charge $Q$ for
various values of $\kappa.$ Black squares correspond to $\kappa=15,$
white squares to $\kappa=10,$ while there follow $\kappa=5,\
\kappa=3$ and $\kappa=1.$ } \label{k_various}
\end{figure}

For the first order numerical solutions of the next section we
used the values of $\kappa$ and $Q$ given in (\ref{values}).
However, to show how the critical temperature depends on the
parameters, in figure \ref{k_various} we show the results of a
numerical study of the critical temperature $T_0$ as a function of
the black hole charge $Q$ for various values of the Einstein
coupling constant $\kappa$, ranging from $\kappa = 1$ to $\kappa =
15$. The critical temperature diverges as $Q\to 0$ (Schwarzschild
limit) for all values of $\kappa$.

\section{First order solution}
\label{sec:3}

At first order, the field equations yield a solution near
the critical temperature, where the scalar field $\varphi$ backreacts on the metric.

To extract the first-order equations from the full set of
non-linear field equations,  notice that the stress-energy tensor
to this order has contributions from the electromagnetic
stress-energy tensor (\ref{3str}), and the scalar field (eqs.\
(\ref{scalarstess}) and (\ref{eqtheta}), with $\varphi$ replaced
by $\varepsilon\varphi^{(0)}$ and $g_{\mu\nu}$ replaced by the RN
solution $g_{\mu\nu}^{(0)}$). For the contribution
(\ref{eqtheta}), after some algebra we obtain the explicit
expression \be \label{eqTh} \Theta_t^t = \Theta_z^z = -
\Theta_\theta^\theta = - \Theta_\phi^\phi = \varepsilon^2
\frac{2z^6 e^{-3\alpha}}{\mu^4 l_0^6 \sinh^2 B} \left[ z^2
{\mathcal{Z}'}^2 \cos^2\theta + \mathcal{Z}^2 \sin^2\theta \right]
+ \mathcal{O} (\varepsilon^4) ~. \ee
%We want to solve the Einstein-Maxwell equations in the full
%dynamical background given by the  total stress-energy tensor
%\[ T_{\mu\nu} = T_{\mu\nu}^{(EM)} +
%T_{\mu\nu}^{(\varphi)} + \kappa \Theta_{\mu\nu}~, \] where all the
%fields are functions of $(r,\theta)$.
 We give the
technical details of the solution of the equations
(\ref{eq1n})-(\ref{eq6n}) in Appendix A.
%\subsection{Numerical Study}
Having the solutions of the equations (\ref{eq1n})-(\ref{eq6n}) we
will numerically determine the metric functions $\alpha$ and
$\beta$ and the electric potential $A$. Notice that the order parameter $\varepsilon$ and the normalization of the zeroth-order scalar field $\varphi^{(0)}$ are not independently defined, since only their product (which ought to be small for the perturbative expansion to be valid) enters the field equations. For convenience, we set $\varepsilon = 1$ in our numerical calculations, making the normalization of $\varphi^{(0)}$ small.

We wand to find the functions $\alpha_{10}(z)$, $A_{10}(z)$,
$\alpha_{12}(z)$, $A_{12}(z)$ which appear in the first order
corrections
\bea \alpha_1 &=& \alpha_{10} (z) P_0(\cos\theta) + \alpha_{12} (z) P_2(\cos\theta) ~, \nonumber\\
A_1 &=& A_{10} (z) P_0(\cos\theta) + A_{12} (z) P_2(\cos\theta)~.
\eea

\begin{figure}
\begin{center}
\includegraphics[scale=0.9,angle=0]{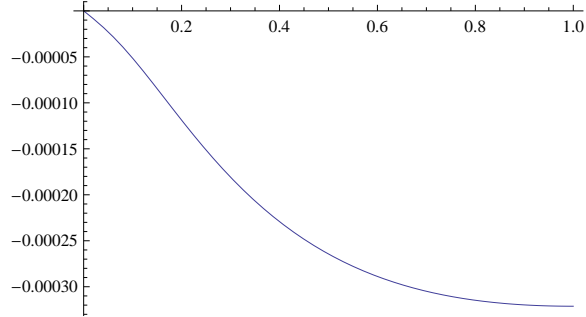}
\end{center}
\caption {The function $\alpha_{10}(z)$ of the angle-independent part of the first-order correction to
the metric.}
\label{da0}
\end{figure}

We will calculate first the angle-independent first-order
corrections. For the correction $\alpha_{10}$ we work with equation
(\ref{zerorder}) noticing that at the boundary $\alpha_{10}(z) \sim
\lambda z,$ while at the horizon such a solution yields an
expression $a + b \ln (1-z).$ Fine tuning of $\lambda$ will yield
$b=0,$ that is, a regular solution. On the technical side we change
the variable $\alpha_{10}$  to $\zeta_0 \equiv
\frac{\alpha_{10}}{z}$ with the new boundary conditions $\zeta_0(0)
= \lambda,\ \zeta_0^\prime(0)=0.$ Figure \ref{da0} depicts the
results for $\alpha_{10}$  for $\lambda = -0.000421852,$ which is
found to yield a regular solution.

\begin{figure}
\begin{center}
\includegraphics[scale=0.9,angle=0]{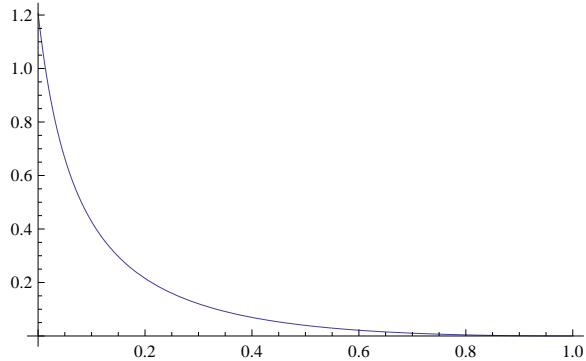}
\end{center}
\caption {The electrostatic potential $A(z)$ at first order.} \label{A0inv}
\end{figure}

The first-order correction to the electric potential $A(z) =
A_0+\varepsilon^2 A_{10}$ may be directly determined if $\alpha$ is
known using (\ref{potential}) and the result is plotted in figure
\ref{A0inv}.\\
A procedure  similar to the one used for the scalar field
will be used for the solution of the equations  (\ref{A2}) and
(\ref{a2}) for the angle-dependent first order corrections $A_{12}$ and
$\alpha_{12}.$ We use the expansions: \be \alpha_{12}(z) = z^3
\sum_{n=1}^{\infty} \tilde{\alpha}_{12}^{( n)} (1-z)^n, \ \ \
A_{12}(z) = z^3 \sum_{n=1}^{\infty} \tilde{A}_{12}^{( n)} (1-z)^n, \label{eq3-10}\ee
where $\tilde{\alpha}_{12}^{(n)},\ \tilde{A}_{12}^{(n)}$ depend on a
single free parameter which can be chosen to be
$\tilde{A}_{12}^{(2)}$ (it is easily seen that the first term
vanishes, $\tilde{A}_{12}^{(1)} = 0$). This parameter can be
determined by a variational method. To this end, multiply
(\ref{A2}) by $A_{12}$ and integrate.
We obtain \be\label{eqA2} \int_0^1 dz \left[ -
(A_{12}')^2 -\left( - \frac{l^2}{z^2D^2} +\frac{2}{z^2D} - \frac{1-3z^2}{z^2 l_0^2} + \frac{6}{z^2} \right) A_{12}^2 + \frac{Ql_0}{\mu D^2} A_{12} \alpha_{12}' \right] = 0 \ee
where we defined
\be D(z) = 1 + z^2 + 2z\coth B~. \ee
Similarly, if we multiply (\ref{a2}) by $\alpha_{12}$, we obtain \be
\int_0^1 dz \left[ - (\alpha_{12}')^2 - \left( - \frac{1+z^2}{l_0^2} + \frac{6}{z^2} +
\frac{8}{(D\sinh B )^2} \right) \alpha_{12}^2
+
\frac{16\pi G Q}{\mu l_0} \alpha_{12}A_{12}' - 16\pi G \frac{\kappa
\mathcal{Q}_2}{z^2} \alpha_2\right] = 0 ~. \label{eqa2}\ee Both
of these equations must be satisfied at the right value of
$\tilde{A}_{12}^{(2)}$. Either one of them determines this value.

\begin{figure}
\begin{center}
\includegraphics[scale=0.9,angle=0]{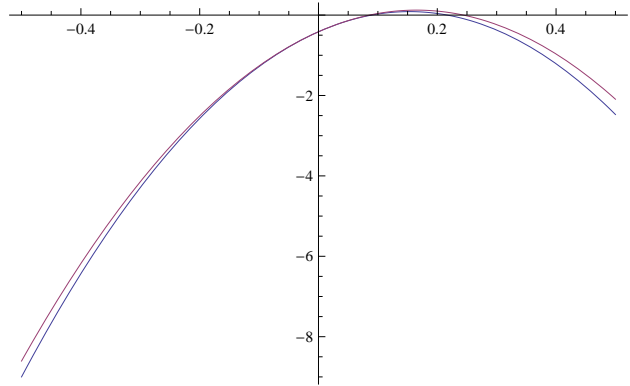}
\end{center}
\caption {Variational expressions (\ref{eqA2}) and (\ref{eqa2}) {\em vs.}\
$\tilde{A}_{12}^{( 2)}$ showing that $\tilde{A}_{12}^{(2)}=0.2.$}
\label{var}
\end{figure}

\begin{figure}
\begin{center}
\includegraphics[scale=0.9,angle=0]{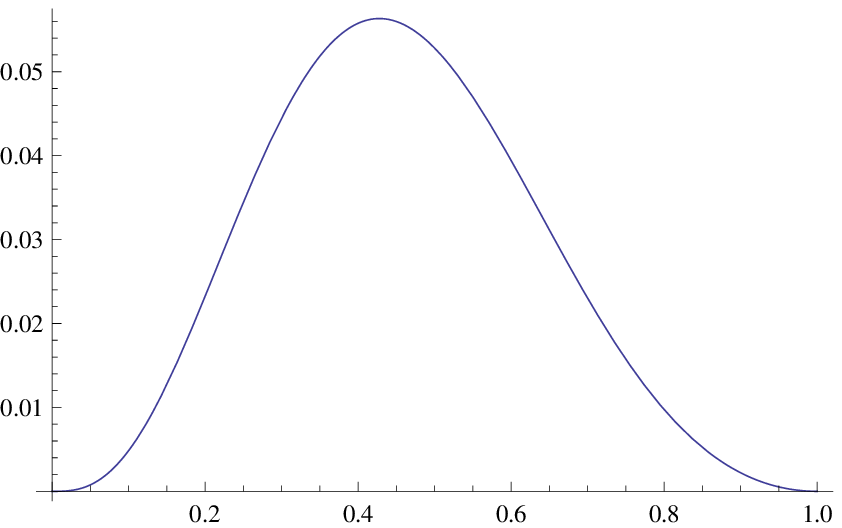}
\end{center}
\caption {The function $A_{12}(z)$ for the parameters of
figure \ref{kappa} and $\tilde{A}_{12}^{(2)}=0.2.$} \label{A}
\end{figure}

\begin{figure}
\begin{center}
\includegraphics[scale=0.9,angle=0]{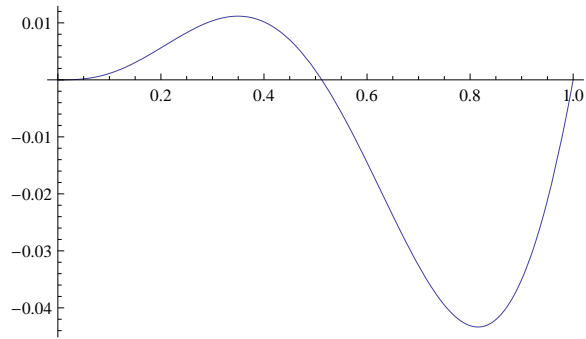}
\end{center}
\caption {The function $\alpha_{12}(z)$ for the parameters
of figure \ref{kappa} and $\tilde{A}_{12}^{(2)}=0.2.$} \label{alpha}
\end{figure}

In figure \ref{var} we show the results for the variational
expressions versus the undetermined parameter
$\tilde{A}_{12}^{(2)}$ and find that $\tilde{A}_{12}^{(2)}=0.2.$
We depict the left-hand side of (\ref{eqa2})  and the sum of
(\ref{eqA2}) and (\ref{eqa2}) keeping four terms in each of the
expansions (\ref{eq3-10}). Notice that the two curves are almost
indistinguishable and both approach zero at the same value,
confirming the consistency of our numerical approach. In figure
\ref{A} we depict the solution for $A_{12},$ while in figure
\ref{alpha} the solution for $\alpha_{12}.$ Therefore, the first
order corrections $A_{12}$ and $\alpha_{12}$ are regular
everywhere.

\begin{figure}
\begin{center}
\includegraphics[scale=0.9,angle=0]{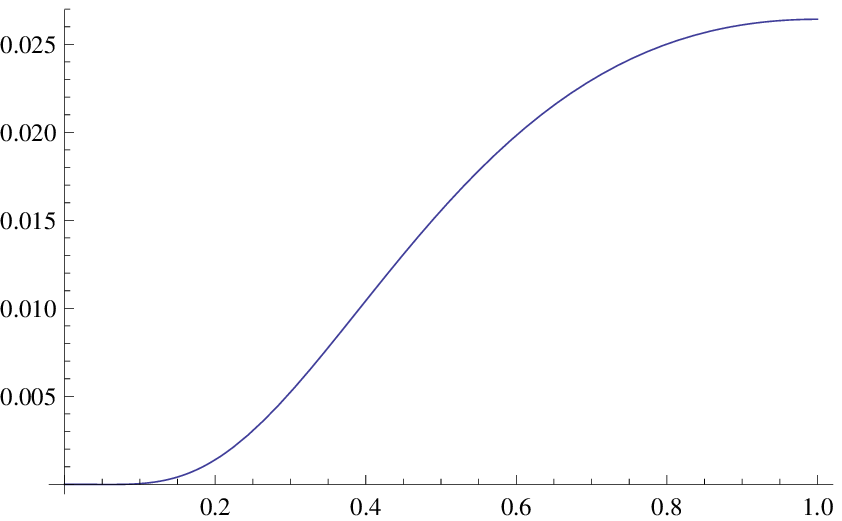}
\end{center}
\caption {The function $\beta_{10}$.} \label{beta0}
\end{figure}

\begin{figure}
\begin{center}
\includegraphics[scale=0.9,angle=0]{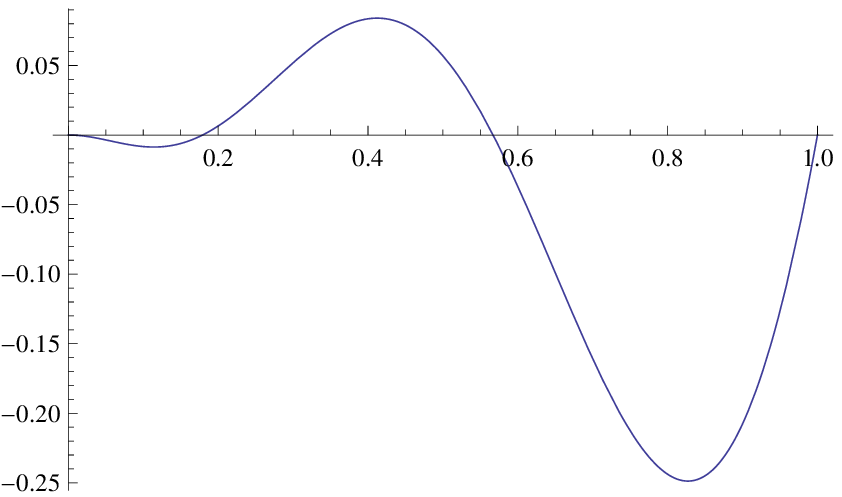}
\end{center}
\caption {The function $\beta_{11}$.} \label{beta1}
\end{figure}

 Finally, the metric function $\beta$
which is given by (\ref{beta}) has as the first order angle-independent correction
the function $\beta_{10}$ of (\ref{beta00a}) which is shown in figure
\ref{beta0}.
%with the boundary condition $\beta_0(0)=0$ .
The function $\beta_1$ in the angle-dependent first-order contribution
given by (\ref{beta11a})
is shown in figure \ref{beta1}. Therefore both corrections are
regular in the entire range of $z$.

%{\bf Can we show the shift of the horizon?}

\section{Thermodynamical stability at first order}
\label{sec:5}

In this section we will discuss the thermodynamical stability of
our solution. We need to know the temperature of the hairy
solution. The temperature of the hairy black hole at first order
expressed in terms of the critical temperature is \be T_{hair} =
T_{0} e^{\varepsilon^2 (-\alpha_{10}(1) + \beta_{10}(1)/2)}~. \ee
Notice that $T_{hair} \ge T_0$, so the RN black hole is unstable
at high temperatures (above $T_0$) for a fixed charge $Q$. As the
mass $M$ approaches its minimum value at extremality, the RN black
hole becomes stable.

Since the correction is quadratic in the scalar field, we deduce
for the value of the scalar field at the horizon, \be \varphi_+
\equiv \varphi\Big|_{z=1} = \gamma \sqrt{ \frac{T_{hair}}{T_0}
-1}~. \ee Let us find an RN black hole at this temperature. We
shall find one with the same charge $Q$.
%, i.e., same thermodynamic potential $F$ as the original RN black hole at temperature $T_{RN}$ (see eq.~(\ref{eqRNGibbs})).
Call the parameters of this black hole $\mu'=\mu+\varepsilon^2 \mu_1$ and
$B' =B+\varepsilon^2 B_1$. Since $\mu /\sinh B =$~const., we have
\[ B_1 = \tanh B \frac{\mu_1 }{ \mu}~. \]
Another relation between $B_1$ and $\mu_1$ is found from
setting
\[ \frac{\delta T}{T_0} = \varepsilon^2 \left[ - \alpha_{10} (1) + \frac{\beta_{10}(1)}{2}
\right]~. \] We obtain
\[  - \alpha_{10} (1) + \frac{\beta_0(1)}{2}  = \frac{4}{e^{2B} -1} B_1 -
\frac{\mu_1}{\mu} ~. \] These two relations determine $B_1$ and $\mu_1$. For the free energy, we deduce
\[ \frac{\delta F_{RN}}{F_{RN}} = \varepsilon^2\frac{1+e^{2B}}{3-e^{2B}} \left[ - \delta\alpha_0 (1) +
\frac{\beta_0(1)}{2} \right] ~. \] Notice that $\delta
F_{RN} > 0$ as long as $e^{2B} < 3$.

We need to compare it with the free energy of the hairy black
hole. The mass of the hairy black hole is found from the
asymptotic behaviour of $\alpha$,
\[ GM = \frac{\mu}{2} \alpha_0'(0)~, \]
where prime denotes differentiation with respect to $z$.
Therefore,
\[ GM_{hair} = GM_{RN} + \varepsilon^2\frac{\mu}{2} \alpha_{10}'(0)~. \]
This is found from the surface (Gibbons-Hawking) term in the
action. The entropy is found from the area of the horizon,
\[ S_{hair} = S_{RN} e^{-\varepsilon^2 [ -\alpha_{10} + \beta_{10}/2 ]} \Big|_{z=1}~. \]
Therefore, the product $TS$ remains unchanged.

There is an additional contribution from the Einstein-Hilbert
action because  the Ricci scalar does not vanish. We obtain a
contribution to the free energy
\[ \delta F_{EH} = -\int d^3 x \sqrt{-g} \frac{R}{16\pi G} = - \frac{\varepsilon^2}{2} \int d^3
x \sqrt{-g} (\partial\varphi^{(0)})^2~. \] This is the dominant
change in the free energy and is clearly negative. Explicitly,
\[  \delta F_{EH} = - \frac{2\pi\mu}{3} \int_0^1 \frac{dz}{z^2} \varepsilon^2\left[
(z\mathcal{Z}')^2 + 2\mathcal{Z}^2 \right] < 0~. \] Therefore, the
difference in free energies is \be \Delta F = \delta F_{hair}
-\delta F_{RN} = \frac{1}{2} \varepsilon^2\alpha_{10}'(0) F_{RN} -
\frac{1+e^{2B}}{3-e^{2B}} \varepsilon^2\left[ - \alpha_{10} (1) +
\frac{\beta_{10}(1)}{2} \right] F_{RN} + \delta
F_{EH}~\label{fren}~. \ee Putting  numbers in (\ref{fren}) we find
$\Delta F=-0.08091$, showing that the hairy black hole is
thermodynamically stable.

\section{Discussion of the solution}
\label{sec:4}

To complete the first-order solution and verify the validity of the perturbative expansion, we determine the first-order correction to the scalar field and calculate various invariants of the metric and show that they are regular at and outside the horizon, showing that no singularity arises at this order.

\begin{figure}
\begin{center}
\includegraphics[scale=0.8,angle=0]{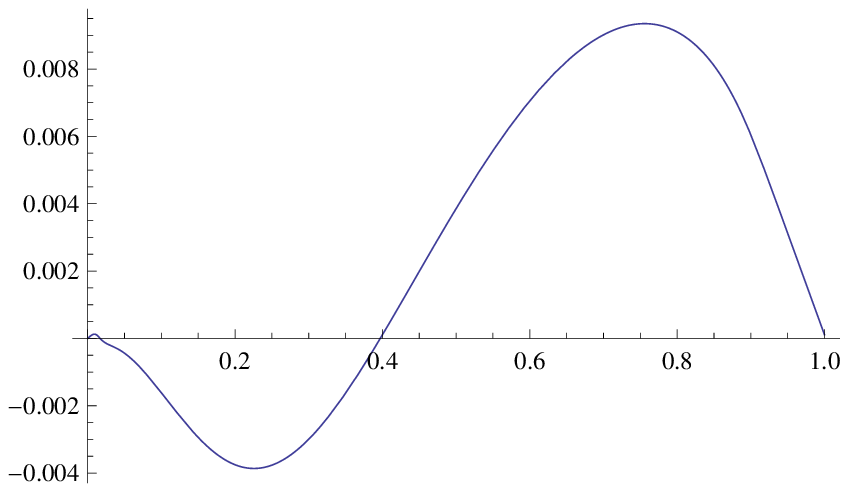}
\includegraphics[scale=0.8,angle=0]{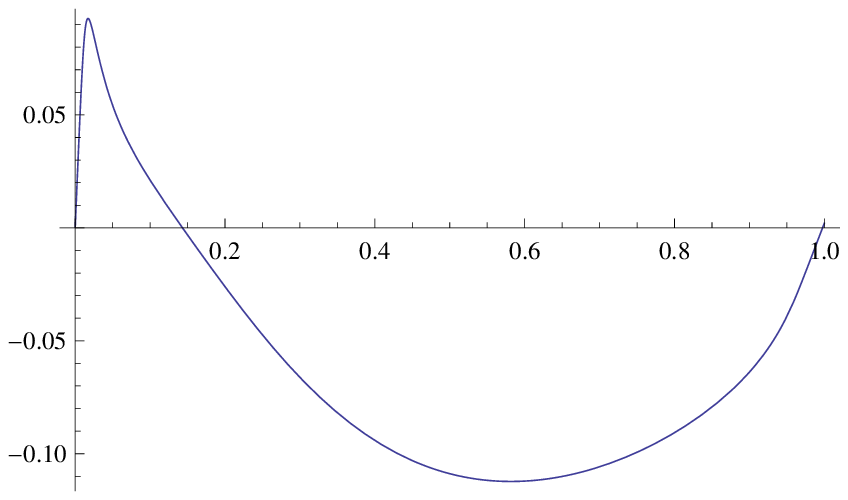}
\end{center}
\caption {The first-order corrections of the scalar field, $\varphi_{10}(z)$ (left panel) and
$\varphi_{11}(z)$ (right panel).}
\label{scalar_first}
\end{figure}

With our choice of the zeroth-order scalar field $\varphi^{(0)}$ as a dipole (eqs.\ (\ref{eqphi0}) and (\ref{eqell1})),
the first-order correction (eq.\ (\ref{pertfull})) contains both a dipole and a $\ell = 3$ term.
Let \be
\varphi^{(1)} (z,\theta) = \varphi_{10}(z) \cos\theta + \varphi_{11}(z) \cos 3
\theta~.\ee The field equation obeyed by $\varphi^{(1)}$ is obtained by collecting the first-order terms in the Klein-Gordon equation  (\ref{kleingordon}). The resulting equation is too long to be included here. It is straightforward to see that it results into decoupled equations for $\varphi_{10}$ and $\varphi_{11}$. The latter involve the functions $\varphi^{(0)} (z), \ \alpha_{10}(z),
\ \alpha_{12}(z), \ \beta_{10}(z)$ and $\beta_{11}(z)$, which have already been calculated. Both $\varphi_{10}(z)$ and
$\varphi_{11}(z)$ can be seen to behave as $c_{10} z^2$ and $c_{11}
z^2$ in the limit $z \rightarrow 0.$ One may tune the coefficients $c_{10}$ and
$c_{11}$ to ensure that the corrections vanish at $z=1.$ The
results for the two functions are depicted in figure
\ref{scalar_first}. It is readily seen, upon comparison with
figure \ref{R}, that the corrections are of the order of the
zeroth order contribution, so the series in $\varepsilon$ is expected
to have a finite radius of convergence.

%\section{Modified Einstein tensor and gauge dependence}

\begin{figure}
\centering
\includegraphics[scale=0.4,angle=0.0]{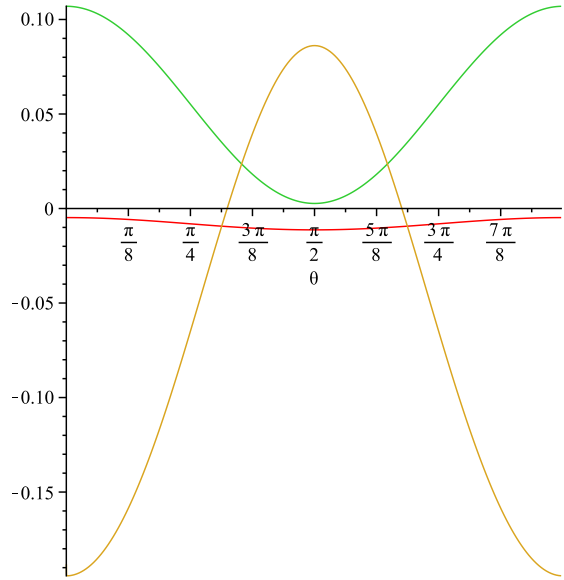}
\caption{The Ricci scalar $R$ \emph{vs.}\ $\theta$ for $z=0.3, \ 0.6,\ 0.9.$ Correspondingly, $R=-0.011,$ $0.0027,$ $0.086$ at $\theta = \frac{\pi}{2}=0.157.$} \label{f1}
\end{figure}

\begin{figure}
\centering
\includegraphics[scale=0.4,angle=0.0]{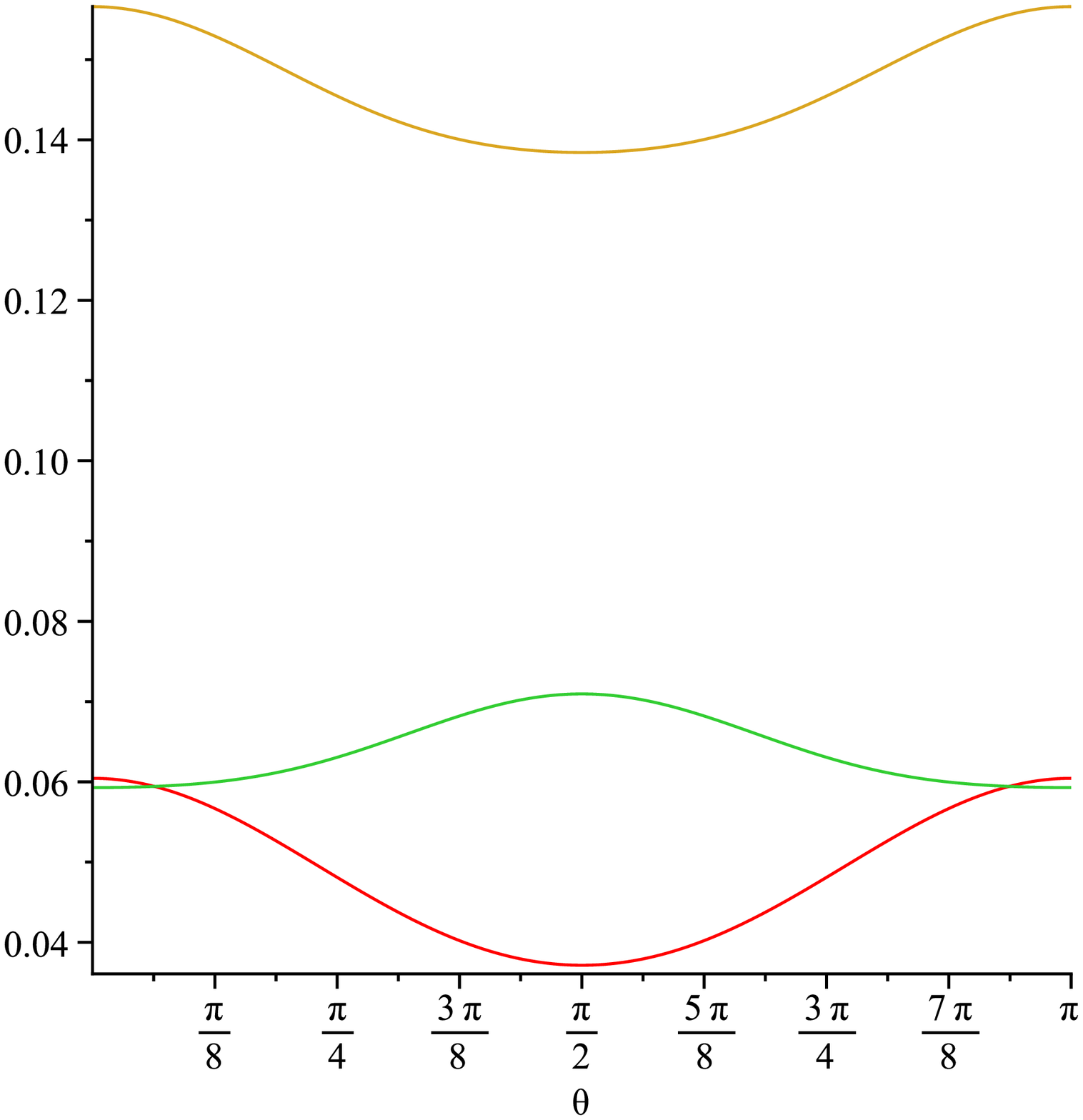}
\includegraphics[scale=0.4,angle=0.0]{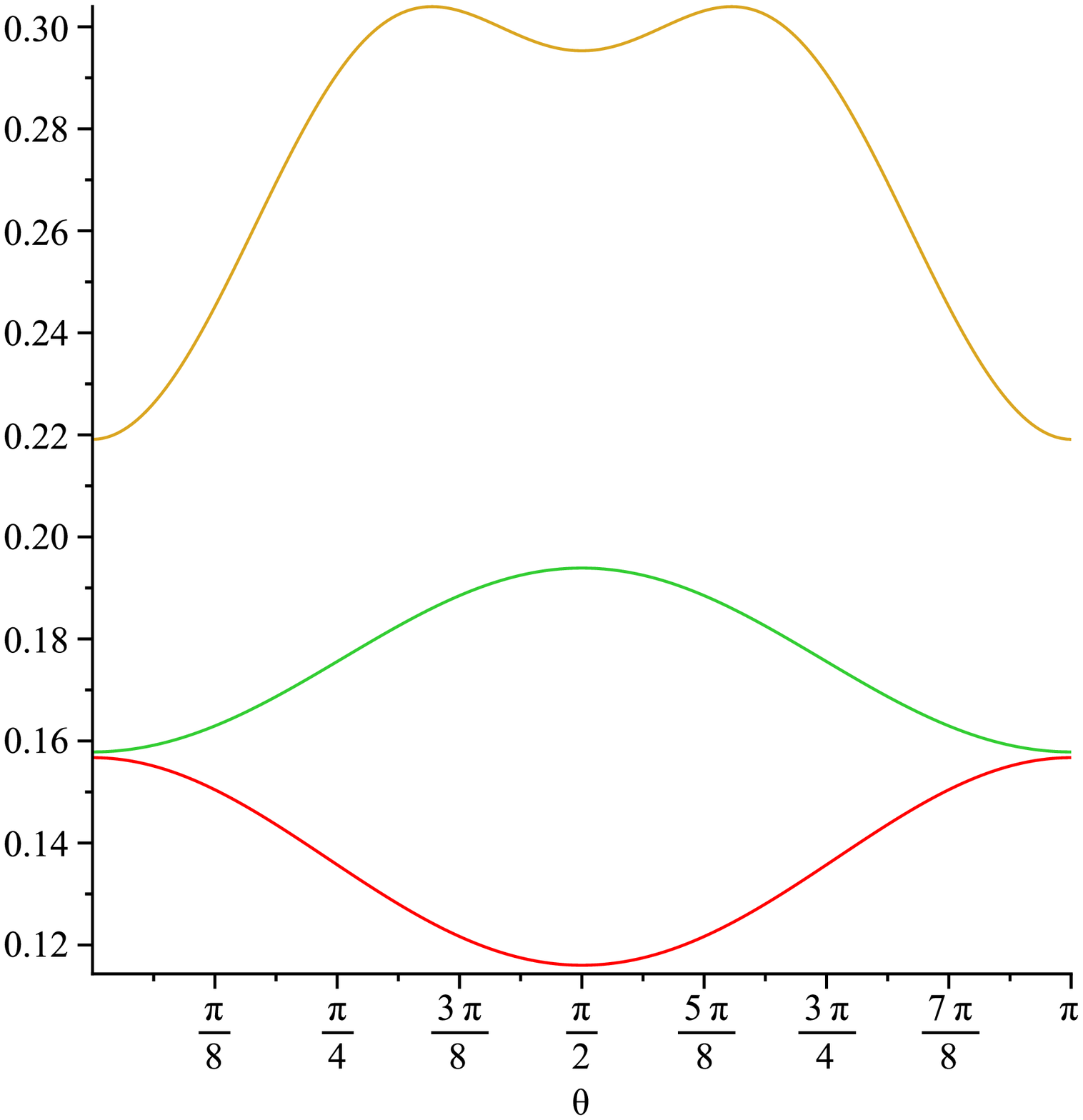}
\caption{$R_{\mu\nu}R^{\mu\nu}$ (left panel)
and $R_{\mu\nu\rho\sigma}R^{\mu\nu\rho\sigma}$ (right panel) \emph{vs.}\ $\theta$ for  for $z=0.3, \ 0.6,\ 0.9$ (bottom to top).}
\label{f2}
\end{figure}

Having demonstrated regularity of the first-order corrections to the
scalar field, we now turn to addressing the same question regarding
the metric. To this end, we need to compute gauge-invariant
quantities, such as the Ricci scalar. The Ricci scalar $R$ vanishes
at zeroth order since it corresponds to a Reissner-Nordstr\"om black
hole. We computed $R$ at first order both analytically and
numerically and found that it is regular everywhere. In figure
\ref{f1} we plot $R$ versus $\theta$ for representative values of
the radius, namely $z=0.3,\ z=0.6$ and $z=0.9.$ Furthermore, we
computed two additional gauge-invariant quantities,
$R_{\mu\nu}R^{\mu\nu}$ and
$R_{\mu\nu\rho\sigma}R^{\mu\nu\rho\sigma}$ at first-order both
analytically and numerically and found them to be regular everywhere
(see figure \ref{f2}). We show the values of the product
$R_{\mu\nu}R^{\mu\nu}$ of the Ricci tensors, as well as the product
$R_{\mu\nu\rho\sigma}R^{\mu\nu\rho\sigma}$ of the Riemann tensors
versus $\theta$ for various values of $z$. We find again that the results are
finite and roughly of the same order of magnitude for the three
typical values of $z.$

\section{Conclusions}
\label{sec:6}

We studied the effect of the presence in the Einstein-Hilbert
action of a derivative coupling of a scalar field to Einstein
tensor, on static black hole solutions. We considered the
Reissner-Nordstr\"om black hole solution in isotropic coordinates
and in this background we introduced a scalar field coupled to
Einstein tensor. For small values of the scalar field we studied
in details how the derivative coupling backreacts on the metric,
solving the full coupled dynamical system of
Einstein-Maxwell-Klein-Gordon equations.

We found that the Reissner-Nordstr\"om black hole above a certain
critical temperature is destabilized to a new hairy black hole
configuration. We studied the properties of this new hairy black
hole solution near the critical temperature and we showed that the
scalar field is regular on the horizon and at infinity. The
no-hair theorem is evaded due to the presence of the derivative
coupling of the scalar field to the Einstein tensor. This new
``Einstein hair" solution is in general anisotropic with the scalar
field and the metric functions to depend also on the angular
coordinate.  We calculated the mass and the temperature of the new
hairy black hole solution and by considering the free energies  we
showed that the new hairy black hole configuration is
thermodynamically stable.

It would be interesting to extend the analysis to more general
``Einstein hair" by including mass and charge for the scalar hair
and explore the existence of spherically symmetric hair. In
particular, it would be of great interest to see if hair can
develop down to zero temperature and form a configuration of
vanishing entropy. Work in this direction is in progress.

\section*{Acknowledgments}

We thank Christos Charmousis for useful discussions.  G.~S.~was
supported in part by the US Department of Energy under grant
DE-FG05-91ER40627. T. K. acknowledges support from the Operational
Program ``Education and Lifelong Learning'' of the National
Strategic Reference Framework (NSRF) - Research Funding Program:
Heracleitus II, co-financed by the European Union (European Social
Fund - ESF) and Greek national funds.

\begin{appendix}

%\appendix{Appendix}

\section{First-order solution of Einstein-Maxwell equations}
In this appendix we give the technical details of solving
the field equations (\ref{eq1n})-(\ref{eq6n}) at first order in the order parameter $\varepsilon^2$. We start with (\ref{eq1n}).
 On the
right-hand side, only $\varphi^{(0)}$ contributes and only through
$\Theta_{\mu\nu}$. Using (\ref{eqTh}), we obtain
\be  z^2 \partial_z^2 l -z\partial_zl + \partial_\theta^2 l + 2\cot\theta
\partial_\theta l = 8\pi G \frac{\mu^2\kappa}{z^2}
l_0^{3} e^{\alpha}
(-\Theta_t^t - \Theta_\phi^\phi ) + \mathcal{O} (\varepsilon^4)= \mathcal{O} (\varepsilon^4)~, \label{aeq1n1}
\ee which has as a solution
\be l = 1 -
z^2 + \mathcal{O} (\varepsilon^4)~, \ee therefore $l$ is unchanged at first order, $l_1=0$ (recall eqs.\ (\ref{eq2-3}) and (\ref{eqz}) for
an RN black hole).

Next we solve the Maxwell equation which takes the explicit form
 \be z^2
\partial_z^2 A + z^2\partial_z(\alpha +
\ln l ) \partial_z A + \partial_\theta^2 A +
\left[ \cot\theta + \partial_\theta \alpha + \partial_\theta \ln l
\right] \partial_\theta A = 0~. \label{eq6ne}\ee
At first
order,
\be \label{eqMn1} z^2 \partial_z^2 A + \left[ -\frac{2z}{l} + \partial_z\alpha
\right] z^2 \partial_z A
+ \partial_\theta^2 A + \cot\theta \partial_\theta A =  \mathcal{O} (\varepsilon^4)~. \ee From
(\ref{eq2n}), at first order we have
\[ z^2 \partial_z^2 \alpha - \frac{2z^3}{l} \partial_z \alpha
+ \partial_\theta^2\alpha + \cot\theta \partial_\theta \alpha =
8\pi G e^\alpha \left[ (z\partial_z A)^2 + 2 \mu^2 \kappa
\frac{l^2}{z^2} \Theta_t^t \right] + \mathcal{O} (\varepsilon^4)~. \] We expand the first-order corrections in Legendre
polynomials,
\bea \alpha_1 &=& \alpha_{10} (z) P_0(\cos\theta) + \alpha_{12} (z) P_2(\cos\theta) ~, \nonumber\\
A_1 &=& A_{10} (z) P_0(\cos\theta) + A_{12} (z) P_2(\cos\theta)~, \nonumber\\
\mu^2 \frac{l^2}{z^2} e^\alpha \Theta_t^t &=& \varepsilon^2 [\mathcal{Q}_0(z) P_0(\cos\theta) +
\mathcal{Q}_2(z) P_2(\cos\theta) ]  + \mathcal{O} (\varepsilon^4)~, \eea
where
\be \mathcal{Q}_0 = \frac{\mathcal{R} }{6 }  \left[ \frac{1}{2} (z\mathcal{Z}')^2 + \mathcal{Z}^2 \right]~,
\ \ \ \
\mathcal{Q}_2 = \frac{\mathcal{R} }{6 } \left[ (z\mathcal{Z}')^2 - \mathcal{Z}^2 \right]~,
\ee
and $\mathcal{R}$ is given in (\ref{eqcalR}).

Using
\[ P_\ell'' (\cos\theta) + \cot\theta P_\ell'(\cos\theta) = - \ell (\ell+1) P_\ell
(\cos\theta)~, \]
we obtain
\begin{eqnarray} (A_0+\varepsilon^2 A_{10})'' + \left[ -\frac{2z}{l} + (\alpha_0+\varepsilon\alpha_{10})' \right] (A_0+\varepsilon^2 A_{10})' &=& \mathcal{O} (\varepsilon^4)
\label{eqx1}~, \\
A_2'' + \left[ -\frac{2z}{l_0} + \alpha_0' \right] A_2' -\frac{6}{z^2} A_2 &=& -
\alpha_2' A_0' \label{eqx2}~, \\
(\alpha_0+\varepsilon^2\alpha_{10})'' - \frac{2z}{l} (\alpha_0+\varepsilon^2\alpha_{10})' - 8\pi G e^{\alpha_0+\varepsilon^2\alpha_{10}} {(A_0+\varepsilon^2 A_{10})'}^2 &=& 16\pi G
\kappa \frac{\varepsilon^2\mathcal{Q}_0}{z^2}  + \mathcal{O} (\varepsilon^4)~,
\nonumber\\
\label{eqx3} \\
\alpha_{12}'' - \frac{2z}{l_0} \alpha_{12}' - \frac{6}{z^2}
\alpha_{12} -8\pi G e^{\alpha_0} {A_0'}^2 \alpha_{12} - 16\pi G
e^{\alpha_0} A_0' A_2' &=& 16\pi G \kappa
\frac{\mathcal{Q}_2}{z^2}~. \label{eqx4}
\end{eqnarray}
From (\ref{eqx1}) we deduce
\be\label{potential} A_0(z) + \varepsilon^2 A_{10} (z) = \frac{Q}{\mu} \int_z^1 \frac{dz'}{l(z')}
e^{-\alpha_0(z') - \varepsilon^2\alpha_{10} (z')}  + \mathcal{O} (\varepsilon^4)~. \ee
Eq.\ (\ref{eqx3}) implies
%\[ \alpha_0 = \alpha_{RN} + \delta\alpha_0~, \]
%where $\alpha_{RN}$ is the Reissner-Nordstr{\o}m solution given in (\ref{eq2-3})  and
that the angle-independent part of first-order correction of $\alpha$ satisfies
\be\label{zerorder} \alpha_{10}'' - \frac{2z}{1-z^2} \alpha_{10}' -
\frac{8}{[(1+z^2)\sinh B + 2z \cosh B ]^2} \alpha_{10} = 16\pi G \frac{\kappa
\mathcal{Q}_0}{z^2}~. \ee
%Notice that there is no dependence on $\mu$ (except through $\kappa/\mu^2$, as before).
To solve it, first we look at the boundary
conditions. At the boundary, $\alpha_{10}'' \approx 0$,
therefore
\be \alpha_{10} \sim \lambda z~. \ee
At the horizon, $(1-z)\alpha_{10}'' - \alpha_{10}' \approx
0$, therefore $\alpha_{10} \sim$const., or $\alpha_{10} \sim
\ln (1-z)$. The solution that asymptotes to $\lambda z$ as $z\to
0$ yields a mixture $\alpha_{10} \sim a + b \ln (1-z)$ at the
horizon. There is a unique value of $\lambda$ for which $b=0$ and
the solution is regular.

The remaining two equations form a coupled system to be solved for
the angle-dependent first-order contributions $A_{12}$ and $\alpha_{12}$. Explicitly,
\begin{eqnarray} A_{12}'' - \frac{2}{z} \left[ \frac{l_0}{1+z^2 + 2 z \coth B}
- \frac{1}{l_0} \right] A_{12}' - \frac{6}{z^2} A_{12} &=& - \frac{Q}{\mu} \frac{l_0}{(1+z^2+2z
\coth B)^2} \alpha_{12}' ~, \nonumber\\ \label{A2} \\
\alpha_{12}'' - \frac{2z}{l_0} \alpha_{12}' - \left[ \frac{6}{z^2} +
\frac{8}{[(1+z^2)\sinh B + 2 z \cosh B]^2} \right] \alpha_{12} &=& -
\frac{16\pi G Q}{\mu} \frac{1}{l_0} A_{12}' + 16\pi G
\frac{\kappa \mathcal{Q}_2}{z^2} ~. \nonumber\\ \label{a2}
\end{eqnarray}
As $z\to 0$, we have
\be A_{12} = \mathcal{A} z^{3} + \dots \ , \ \ \alpha_{12} = \mathbf{a} z^3 + \dots \ee
The constants $\mathcal{A}$ and $\mathbf{a}$ are fixed by
demanding $\alpha_{12} = A_{12} = 0$ at the horizon (so there is no
angular dependence of the temperature or an electric field along
the horizon).

Finally, we obtain $\beta$ from the third Einstein equation (\ref{eq3n}).
At zeroth order, we have
\be 0 = -\frac{1}{2} (z\alpha_{0}')^2 -3 \frac{zl_0'}{l_0} + 2 \left( \frac{zl_0'}{l_0}
\right)^2 - \frac{z^2 l_0''}{l_0} + 8\pi G e^{\alpha_{0}} (zA_{0}')^2 ~, \ee
which is easily seen to be satisfied.

To solve the equation at first order, we set
\be\label{beta} \beta_1 = \beta_{10}(z) + \beta_{11}(z) \cos^2\theta~. \ee
and obtain
\be\label{beta00} (1+z^2) z\beta_{10}' =
\mathcal{S}_0 (z)~, \ee where \bea \mathcal{S}_0 (z) &=&
\frac{4z^2[(1+z^2)\cosh B + 2z \sinh B]}{(1+z^2)\sinh B + 2z \cosh
B} \left( \alpha_{10}' - \frac{1}{2} \alpha_{12}' \right)\nonumber
\\ &-& \frac{8z^2(1-z^2)}{[(1+z^2)\sinh B + 2z\cosh B]^2} \left(
\alpha_{10} - \frac{1}{2} \alpha_{12} \right)
\nonumber \\
 &-& 8\pi G (1-z^2) \left[
1 - \frac{4\kappa z^4}{ \mu^2 [ (1+z^2)\sinh B + 2z\cosh B]^4 }
\right] \mathcal{Z}^2~. \eea
to be solved for $\beta_{10} (z)$, and
\be\label{beta11} (1+z^2) z\beta_{11}' -2(1-z^2) \beta_{11} = \mathcal{S}_1 (z)~, \ee
where \bea \mathcal{S}_1 (z) &=& \frac{12 z^2[(1+z^2)\cosh B + 2z
\sinh B]}{(1+z^2)\sinh B + 2z \cosh B} \alpha_{12}' +
\frac{12z^2(1-z^2)}{[(1+z^2)\sinh B + 2z\cosh B]^2}
\alpha_{12}\nonumber \\ &+& 8\pi G (1-z^2) [ (z\mathcal{Z}')^2 + \mathcal{Z}^2 ]
 + 8\pi G
\frac{4\kappa z^4(1-z^2)}{ \mu^2 [ (1+z^2)\sinh B + 2z\cosh B]^4 }
[ (z\mathcal{Z}')^2 - \mathcal{Z}^2 ]~, \nonumber \\
\eea
to be solved for $\beta_{11}(z)$.

They are both first-order equations and are easily integrated. We obtain
\be\label{beta00a} \beta_{10} (z) = \int_0^z \frac{dy}{y(1+y^2)}\, \mathcal{S}_0 (y) \ee
where we used the boundary condition $\beta_{10} (0) = 0$.

For $\beta_{11}$, we need $\beta_{11}=0$ at both ends. We obtain
\be\label{beta11a} \beta_{11} = -\frac{z^2}{(1+z^2)^2} \int_z^1 \frac{dy}{y^3} (1+y^2)
\mathcal{S}_1 (y) \ee
where the limit of integration was chosen so that $\beta_{11}=0$ at the horizon
($z=1$). Notice that $\beta_{11}\sim \mathcal{O} (z^2)$ as $z\to 0$ ($r\to\infty$), so
the other boundary condition is also satisfied.

\end{appendix}

\end{document}